\documentclass[fleqn,usenatbib]{mnras}

\usepackage{newtxtext,newtxmath}
\usepackage{lscape}

\usepackage[T1]{fontenc}
\usepackage{multirow}
\DeclareRobustCommand{\VAN}[3]{#2}
\let\VANthebibliography\thebibliography
\def\thebibliography{\DeclareRobustCommand{\VAN}[3]{##3}\VANthebibliography}

\usepackage{graphicx}	
\usepackage{amsmath}	
\title[Beyond - $\Lambda$CDM constraints from BOSS+eBOSS]{Beyond - $\Lambda$CDM constraints from the full shape clustering measurements from BOSS and eBOSS}

\author[A. Semenaite et al.]{
Agne Semenaite$^{1}$\thanks{E-mail: agne@mpe.mpg.de},
Ariel G. S\'anchez$^{1,2}$,
Andrea Pezzotta$^{1}$,
Jiamin Hou$^{1,3}$,
Alexander Eggemeier$^{4}$,
\newauthor
Martin Crocce$^{5,6}$, 
Cheng Zhao$^{7}$,
Joel R. Brownstein$^{8}$,
Graziano Rossi$^{9}$
and Donald P. Schneider$^{10, 11}$
\\
\\
$^{1}$ Max-Planck-Institut f\"ur extraterrestrische Physik, Postfach 1312, Giessenbachstr., 85748 Garching, Germany\\ 
$^{2}$Universit\"as-Sternwarte M\"uchen, Scheinerstrasse 1, D-81679 M\"uchen, Germany\\
$^{3}$Department of Astronomy, University of Florida,
211 Bryant Space Science Center, Gainesville, FL 32611, USA\\
$^{4}$ Argelander Fellow, Argelander Institut für Astronomie der Universität Bonn, Auf dem Hügel 71, 53121 Bonn, Germany\\
$^{5}$  Institute of Space Sciences (ICE, CSIC), Campus UAB, Carrer de Can Magrans, s/n, 08193 Barcelona, Spain\\
$^{6}$  Institut d’Estudis Espacials de Catalunya (IEEC), 08034 Barcelona, Spain\\
$^{7}$Institute of Physics, Laboratory of Astrophysics, \'Ecole Polytechnique F\'ed\'erale de Lausanne (EPFL), Observatoire de Sauverny, CH-1290 Versoix, Switzerland\\
$^{8}$ Department of Physics and Astronomy, University of Utah, 115 S. 1400 E., Salt Lake City, UT 84112, USA\\
$^{9}$ Department of Physics and Astronomy, Sejong University, Seoul, 143-747, Republic of Korea\\
$^{10}$ Department of Astronomy and Astrophysics, The Pennsylvania State University, University Park, PA 16802\\
$^{11}$ Institute for Gravitation and the Cosmos, The Pennsylvania State University, University Park, PA 16802
}

\date{Accepted XXX. Received YYY; in original form ZZZ}

\pubyear{2015}

\begin{document}
\label{firstpage}
\pagerange{\pageref{firstpage}--\pageref{lastpage}}
\maketitle

\begin{abstract}
We analyse the full shape of anisotropic clustering measurements from the extended Baryon Oscillation Spectroscopic survey (eBOSS) quasar sample together with the combined galaxy sample from the Baryon Oscillation Spectroscopic Survey (BOSS). We obtain constraints on the cosmological parameters independent of the Hubble parameter $h$ for the extensions of the $\Lambda$CDM models, focusing on cosmologies with free dark energy equation of state parameter $w$. We combine the clustering constraints with those from the latest CMB data from Planck to obtain joint constraints for these cosmologies for $w$ and the additional extension parameters - its time evolution $w_{\rm{a}}$, the physical curvature density $\omega_{K}$ and the neutrino mass sum $\sum m_{\nu}$. Our joint constraints are consistent with flat $\Lambda$CDM cosmological model within 68\% confidence limits. We demonstrate that the Planck data are able to place tight constraints on the clustering amplitude today, $\sigma_{12}$, in cosmologies with varying $w$ and present the first constraints for the clustering amplitude for such cosmologies, which is found to be slightly higher than the $\Lambda$CDM value. Additionally, we show that when we vary $w$ and allow for non-flat cosmologies and the physical curvature density is used, Planck prefers a curved universe at $4\sigma$ significance, which is $\sim2\sigma$ higher than when using the relative curvature density $\Omega_{\rm{K}}$. Finally, when $w$ is varied freely, clustering provides only a modest improvement (of 0.021 eV) on the upper limit of $\sum m_{\nu}$.
\end{abstract}




\section{Introduction}

The standard cosmological model, $\Lambda$CDM predicts a spatially flat Universe, which today is dominated by cold dark matter (CDM) and dark energy components, the latter believed to be well described by a constant negative equation of state parameter $w=-1$, equivalent to cosmological constant $\Lambda$. This model is supported by a number of cosmological observations spanning a range of redshifts, from Cosmic Microwave Background power spectra probing the epoch of recombination  \citep{wmap9, Planck2018}, to baryonic acoustic oscillation (BAO) feature in galaxy clustering measurements reaching matter dominated redshifts \citep{Eisenstein2005, Cole2005, Anderson2012, Alam17, eboss2021}, to standard candle Supernovae Ia (SN) observations that reveal the dominance of dark energy today \citep{Perlmutter1999, Riess1998}. 

Despite this success, the increasingly precise low-redshift measurements have started hinting at discrepancies between the observed $\Lambda$CDM parameter values and CMB predictions. Most significantly, the SN measurements of the Hubble parameter ($H_0$) suggest an expansion rate of the Universe today that is 5$\sigma$ above the prediction coming from Planck satellite CMB observations \citep{Planck2018, Riess2021}. Additionally, there are also inconsistencies surrounding the amplitude of the weak lensing signal, as described by $S_8=\sigma_8\sqrt{\Omega_{\rm{m}}/0.3}$ (where $\Omega_{\rm{m}}$ is the relative matter density parameter, and $\sigma_8$ is the linear density field variance as measured on a scale of 8 $h^{-1}$Mpc), with different weak lensing surveys finding a value that is 2-3$\sigma$ below Planck's best fit prediction \citep{kids1000, desyr3}.

Galaxy clustering has provided the most precise cosmological low-redshift constraints to date \citep{eboss2021} and is, therefore, crucial in determining whether the discrepancies observed point to a tension between low and high-redshift measurements or are related to specific probes instead. Two-point clustering statistics display two main features that allow us to fit for the background expansion of the Universe and the growth of structure separately. First, the acoustic waves of the baryon-photon plasma in the early pre-recombination Universe appear as the BAO signature - a bump in two-point correlation function, or a series of oscillations in Fourier space, whose physical size is known and can, therefore, be used as a standard ruler to probe the distance--redshift relation. Second, each galaxy has a peculiar velocity component due to the coherent motion towards overdense regions, which provides an additional contribution to the measured redshift and results in an anisotropic clustering pattern - the redshift space distortions (RSD). In two-point statistics, this manifests as an angle-dependent change in amplitude, which can be fit to extract cosmology based on the structure growth. Traditionally, the most common approach has been to model and fit the full shape of the clustering measurements \citep[e.g.,][]{Percival2002,Tegmark2004,Sanchez2006, Sanchez2012, sanchez17,Parkinson2012}. 
This ``full-shape'' analysis has regained attention recently as it makes use 
of all of the information available on the statistic being considered 
\citep{Tr_ster_2020, dAmico_2020, Ivanov_2020, Chen2021}. 
While some of these analyses did find a lower value of $\sigma_8$ \citep{Tr_ster_2020, Chen2021}, overall, no significant discrepancies with Planck have been reported for clustering constraints alone.

A major issue with the discussion surrounding the consistency between large-scale structure probes and Planck is that it has largely been based on the parameters $S_8$ and $\sigma_8$. Though well measured by weak lensing, the two parameters are defined through $h$ (through the variance scale 8 $h^{-1}$ Mpc for $\sigma_8$ and the definition of critical density in $\Omega_{\rm{m}}$) and, therefore, depend on the posterior of $h$ recovered by a particular probe. As noted in \cite{littleh}, this approach leads to inconsistencies between different measurements, which make a meaningful comparison non-trivial.

A recent full shape analysis of Baryon Oscillation Spectroscopic Survey (BOSS) galaxy and extended Baryon Oscillation Spectroscopic Survey (eBOSS) quasar clustering measurements by \cite{bosseboss_lcdm} has addressed this inconsistency by exploring $\Lambda$CDM constraints on the parameter space that is explicitly chosen to not be defined through $h$. Here $\sigma_8$ is replaced by its equivalent defined on a physical scale of 12 Mpc, $\sigma_{12}$, as introduced by \cite{littleh}, and the relative densities $\Omega$ are substituted by their physical counterparts $\omega=\Omega h^2$. In this alternative parameter space the joint clustering dataset BOSS+eBOSS shows a near-perfect consistency with the Planck results. Moreover, even when clustering is further combined with weak lensing $3~\times~2$pt measurements from Dark Energy Survey Year 1 data release \citep[DES Y1, ][]{Abbott18}, the recovered value of $\sigma_{12}$ is in an excellent agreement with the CMB predicted clustering amplitude. 

The tension between weak lensing (and its combination with clustering) and Planck, however, does not disappear completely even in the physical parameter space but is instead reflected in the  $\log(10^{10}A_{\rm{s}})-\sigma_{12}$ plane which relates initial ($A_{\rm{s}}$) and final ($\sigma_{12}$) amplitudes of density fluctuations. For a given value of $\sigma_{12}$, Planck's preferred value of $\log(10^{10}A_{\rm{s}})$ is lower than that of DES or DES+BOSS+eBOSS, indicating a greater predicted amount of structure growth than what is observed by the large scale structure. Furthermore, when the parameters describing the shape of the power spectrum (spectral index $n_{\rm{s}}$ and baryon and cold dark matter densities $\omega_{\rm{b}}$ and $\omega_{\rm{m}}$) are fixed to their Planck best fit values, there is also a slight discrepancy in the recovered dark energy density $\omega_{\rm{DE}}$, with Planck preferring a 1.7$\sigma$ lower value than DES+BOSS+eBOSS. 

While these differences are not yet significant, they are consistent with the amount of tension seen between DES YR1 and Planck \citep{Abbott18} and offer a straightforward interpretation within $\Lambda$CDM relating differences in structure growth with the poorly constrained $\omega_{\rm{DE}}$, whose dominance emerges at low redshift only. As Planck measures CMB at the epoch of recombination ($z_{\star}$), its predictions for low redshifts, set by the angular diameter distance $D_{\rm{A}}(z_{\star})$,  are extremely sensitive to the model choice. Galaxy clustering, conversely, is directly sensitive to dark energy both through the BAO peak (the line-of-sight measurement allows to constrain the expansion rate $H_0$ which is just a sum of physical matter and dark energy densities) and the RSD effect (dark energy slowing down the structure growth). Nevertheless, the current clustering constraints on $\omega_{\rm{DE}}$ are at the level of around 6\% for a fixed dark energy equation of state parameter $w=-1$, and degrade if $w$ is allowed to vary. Nonetheless, the future Stage IV galaxy surveys, such as the Dark Energy Spectroscopic Instrument \citep[DESI, ][]{desi}, the ESA space mission \textit{Euclid} \citep{euclid}, and the Legacy Survey of Space and Time (LSST) at the Rubin Observatory \citep{lsst} promise further improved precision on cosmological parameters, including sub-percent measurements of the fraction of dark energy.

Until these new data become available and in light of the inconsistencies within $\Lambda$CDM among the different probes, a number of extensions to the base $\Lambda$CDM model may be considered, as has already become standard in many major surveys \citep{Spergel_2003, sdss_sn, cfhtlens, desy1_ext, eboss2021, kids_beyond, Planck2018, desy3_ext}. While the constraints derived from BAO and RSD summary statistics have mostly focused on combining clustering with CMB or supernovae data, full shape galaxy clustering analyses have been shown to provide competitive results beyond $\Lambda$CDM without requiring combination with any additional probes \citep{boss_eft}. Importantly, unlike when using these summary statistics, full shape analyses are not susceptible to the bias due to the  $h^{-1}$Mpc units (RSD effect is usually summarised as a combination of linear growth rate $f$ and $\sigma_8$).

 With this motivation in mind, this analysis extends the full shape BOSS galaxy and eBOSS quasar clustering analysis of \cite{bosseboss_lcdm} and presents the current physical parameter space constraints for extensions to $\Lambda$CDM. In particular, we are interested in models where $w$ is allowed to take values other than $w=-1$, and in the resulting constraints for curvature, neutrino mass, and time-varying equation of state parameter in such cosmologies. 

Our data and modelling choices remain largely the same as that of \cite{bosseboss_lcdm} (however, we do not explore combinations with weak lensing, but rather make use of supernovae measurements, as they provide significant additional constraining power at low redshifts, important for evolving dark energy models) and are described in Section \ref{section:methodology}, which also reviews our physical parameter space and priors used. The results for each of the cosmologies considered are presented in Section \ref{section:results}, together with constraints on selected parameters in Table \ref{tab:constraints} (additional parameter constraints are presented in Appendix \ref{appx:A}). We make use of the simplest extension considered, $w$CDM, to illustrate the advantages of physical parameter space in such extended models and discuss this in Section \ref{section:wcdm}. Our conclusions are presented in Section \ref{section:conclusions}.

\section{Methodology}
\label{section:methodology}

This analysis is a follow up on the work by \cite{bosseboss_lcdm} and uses the same clustering measurements and modelling choices. In this section we, therefore, only provide a summary of the main points, referring the reader to \cite{bosseboss_lcdm} and the references therein for a more detailed description. 

\subsection{Galaxy and QSO clustering measurements}

We consider configuration space clustering measurements from BOSS galaxy samples \citep{boss1} and eBOSS quasar (QSO) catalogue \citep{eboss_qso, eboss1}, which are part of Sloan Digital Survey Data Release 12 \citep[SDSS DR12,][]{sdssdr12_1, sdssdr12_2} and Data Release 16 \citep[SDSS DR16,][]{sdssdr16_1,sdssdr16_2}, respectively. Each of the two data vectors represent a measurement of the anisotropic two point correlation function $\xi(s, \mu)$, where $s$ is the comoving galaxy separation and $\mu$ is the cosine of the angle between the separation vector and the line of sight, compressed into either clustering wedges (for BOSS galaxies) or Legendre multipoles (eBOSS QSO). The two statistics carry equivalent information and clustering wedges may be expressed as a linear combination of multipoles.

We consider BOSS galaxy clustering wedges measurements from 
\cite{sanchez17} based on the combined galaxy sample \citep{sdssdr12_2} in two redshift bins: 0$.2<z<0.5$ (corresponding to an effective redshift $z_{\rm{eff}}=0.38$) and $0.5<z<0.75$ ($z_{\rm{eff}}=0.61$). The clustering wedges statistic \citep*{kazin_wedges} is defined as the average of $\xi(s, \mu)$ over an angular interval $\Delta \mu=\mu_2-\mu_1$. Here the clustering wedges are measured in three equal width intervals covering $\mu$ range 0 to 1. The covariance matrices for these data are estimated from the set of 2045 {\sc MD-Patchy} mock catalogues \citep{Kitaura2016}.

For the clustering of eBOSS QSO sample we use Legendre multipole measurements of \cite{Hou2021}. Here $\xi(s, \mu)$ is expressed in Legendre polynomial basis and we consider the non-vanishing multipoles $\ell=0,2,4$. The QSO sample covers the redshift range of $0.8<z<2.2$, with $z_{\rm{eff}}=1.48$ and the covariance matrix is obtained using a set of 1000 mock catalogues, as described in \cite{Zhao2021}.

The treatment of BOSS and eBOSS QSO clustering measurements matches the original treatment by \cite{sanchez17} and \cite{Hou2021}: we consider the scales of $20\,h^{-1}{\rm Mpc} < s< 160\,h^{-1}{\rm Mpc}$ and assume a Gaussian likelihood for each set of measurements, which are taken to be independent (as they do not overlap in redshift). The covariances are kept fixed and we account for the finite number of mock catalogues used in their derivation \citep{kaufman1967,Hartlap2007,Percival2014}. 

\subsection{Model}
\label{sec:model}

The model for the full shape galaxy clustering wedges and Legendre multipoles used in this analysis is identical to that of \cite{bosseboss_lcdm}: we use a response function based approach for the non-linear matter power spectrum  predictions \citep[as in the original eBOSS QSO analysis by][]{Hou2021}. We model galaxy bias at one loop and make use of co-evolution relations for the tidal bias parameters \citep[motivated by the findings of][]{Eggemeier2021} and account for RSD and Alcock-Paczynski (AP) distortions \citep{Alcock1979} as described in the original wedges analysis in \cite{sanchez17}. We also correct for the non-negligible redshift errors in the QSO sample \citep{Zarrouk2018} following \cite{Hou_2018} and including a damping factor to the power spectrum, $\exp\left(-k\mu \sigma_{\rm err}\right)$, with $\sigma_{\rm{err}}$ as a free parameter. 

Our model predictions for the non-linear matter power spectrum $P_{\rm{mm}}(k)$ (the Fourier space equivalent of $\xi(s)$) are obtained using the Rapid and Efficient SPectrum calculation based on RESponSe functiOn approach \citep[{\sc respresso},][]{respresso}. The key ingredient here is the response function, which quantifies the variation of the non-linear matter power spectrum at scale $k$ induced by a change in linear power spectrum at scale $q$. Given a fiducial measurement of $P_{\rm{mm}}(k|\pmb{\theta}_{\rm{fid}})$ from a set of N-body simulations with cosmological parameters $\pmb{\theta}_{\rm{fid}}$, the response function allows one to obtain a prediction for $P_{\rm{mm}}(k)$ for an arbitrary cosmology with multi-step reconstruction used for cosmologies that differ considerably from the fiducial one. In {\sc respresso}, $\pmb{\theta}_{\rm{fid}}$ corresponds to the best-fitting $\Lambda$CDM model to the Planck 2015 data \citep{Planck2015}. The response function is modelled using the phenomenological model of \cite{NisBerTar1712} based on renormalised perturbation theory \citep{regpt}. 

Our bias model follows \cite{Eggemeier2019} and relates the matter density fluctuations $\delta_{\rm{m}}$ to the galaxy density fluctuations $\delta$ at one loop:
\begin{equation}
    \delta = b_1\delta_m+\frac{b_2}{2}\delta_m^2+\gamma_2\mathcal{G}_2(\Phi_v)+\gamma_{21}\mathcal{G}_2(\varphi_1, \varphi_2)+... 
\end{equation}
Here the Galileon operators $\mathcal{G}_2$ of the normalized velocity potential $\Phi_{\nu}$ and linear and second-order Lagrangian perturbation potentials $\varphi_1$ and $\varphi_2$ are defined as:
\begin{align}
\mathcal{G}_2(\Phi_{\nu})&=(\nabla_{ij}\Phi_{\nu})^2 - (\nabla^2\Phi_{\nu})^2,\\
\mathcal{G}_2(\varphi_1, \varphi_2)&=\nabla_{ij}\varphi_2\nabla_{ij}\varphi_1 - \nabla^2\varphi_{2}\nabla^2\varphi_{1}.
\end{align}
In order to reduce the number of free parameters, we express the tidal bias parameters $\gamma_2$ and $\gamma_{21}$ in terms of linear bias $b_1$, as:
\begin{align}
\gamma_{2} &= 0.524-0.547b_1+0.046b_1^2,\\
\gamma_{21}&= - \frac{2}{21}(b_1-1)+\frac{6}{7}\gamma_2.
\label{eq:gamma2}
\end{align}
Here the relation for $\gamma_2$ is as obtained by \cite{tidal1} using excursion set theory, while $\gamma_{21}$ is set by assuming conserved evolution of tracers after their formation \citep{Fry9604,CatLucMat9807,CatPorKam0011,Chan2012}. Using these relations the only free parameters in the bias model are linear and quadratic bias parameters $b_1$ and $b_2$. \cite{pezzotta} tested these relations together with non-linear matter power spectrum prescription from \textsc{respresso} and demonstrated that for BOSS-like samples this approach returns unbiased cosmological constraints.  

Our RSD description follows \cite{RSD} and \cite{Taruya_2010} with the two dimensional redshift-space power spectrum written as a product of the `no-virial' power spectrum, $P_{\rm novir}(k,\mu)$, and a non-linear correction due to fingers-of-God or galaxy virial motions $W_{\infty}(\lambda=ifk\mu)$:
\begin{equation}
P(k,\mu) = W_\infty(i f k \mu) \, P_{\rm novir}(k,\mu).
\label{Prsd}
\end{equation}
$P_{\rm novir}(k,\mu)$ is computed using a one-loop approximation which consists of a term that corresponds to the non-linear version of the Kaiser formula \citep{Kaiser} and two higher-order terms which account for cross- and bispectrum contributions from density and velocity fields. The corresponding velocity-velocity and matter-velocity power spectra are obtained from empirical relations measured from N-body simulations \citep{Bel2019}. Finally, the virial correction $W_{\infty}$ is parametrised as \citep{sanchez17}:
\begin{equation}
W_{\infty}(\lambda)=\frac{1}{\sqrt{1-\lambda^2a^2_{\rm{vir}}}}\,\exp\left(\frac{\lambda^2\sigma^2_v}{1-\lambda^2\mathnormal{a}^2_{\rm{vir}}}\right).
\end{equation}
$W_{\infty}$ is characterised by a single free parameter $a_{\rm{vir}}$ which describes the kurtosis of the small-scale velocity distribution. The one-dimensional linear velocity dispersion $\sigma_{\rm{v}}$ is defined in terms of the linear matter power spectrum $P_{\rm{L}}$:
\begin{equation}
 \sigma_v^2 \equiv \frac{1}{6\pi^2}\int {\rm d}k\,P_{\rm L}(k).
\label{eq:sigmav}
\end{equation}

We account for AP distortions due to the difference between true and fiducial cosmologies by introducing the geometric distortion factors. The line-of-sight distortion ($q_{\parallel}$) is characterised by the ratio of Hubble parameters evaluated at the effective redshift in the fiducial ($H'(z_{\rm{eff}})$) and true ($H(z_{\rm{eff}})$) cosmologies. Equivalently, the distortions perpendicular to the line of sight ($q_{\perp}$) are described by the ratio of comoving angular diameter distances $D_{\rm{M}}(z)$: 
\begin{align}
q_{\bot} &=D_{\rm{M}}(z_{\rm eff})/D_{\rm{M}}'(z_{\rm eff}),\\
q_{\parallel} &=H'(z_{\rm eff})/H(z_{\rm eff}). 
\end{align}
The distortion factors are then used to rescale angles $\mu$ and galaxy separations $s$ such that:
\begin{align}
s &=s'\left( q_{\parallel}^2\mu'^2+q^2_{\bot}(1-\mu'^2)\right),\\
\mu &=\mu'\frac{q_\parallel}{\sqrt{q_{\parallel}^2\mu'^2+q^2_{\bot}(1-\mu'^2)}}.
\end{align}

Our model, therefore, has a total of four free parameters (three for BOSS): $b_1, b_2, a_{\rm{vir}}$ and $\sigma_{\rm{err}}$. For more details and model validation, see \cite{bosseboss_lcdm}. While our model was tested on mocks that correspond to $\Lambda$CDM cosmologies, we expect the results of these tests to be applicable to the extended models as well. The cosmology extensions that we consider (with the exception of massive neutrinos) have the effect of additionally rescaling the amplitude of the matter power spectrum with the final result equivalent to a power spectrum of a $\Lambda$CDM cosmology with an adjusted amplitude, $\sigma_{12}$, as discussed in \cite{evolution_mapping}.

\subsection{Additional data}

We explore the consistency of our clustering data with the CMB temperature and polarization power spectra from {\it Planck} satellite \citep[Planck, ][]{Planck2018}. We use public nuisance parameter-marginalised likelihood \texttt{plik\_lite\_TTTEEE+lowl+lowE} and do not include CMB lensing information. 
In addition to these constraints, we also supplement clustering information with that from Supernovae Ia from the combined `Pantheon' sample \citep{Pantheon18}, consisting of measurements from 1,048 SNe Ia in the redshift range of $0.01 < z < 2.3$. We obtain supernovae cosmological constraints from the JLA likelihood module as implemented in \textsc{CosmoMC} \citep{Lewis_2002}. 

\subsection{Parameters and priors}

\begin{table}
	\centering
	\caption{Priors used in our analysis. $U$ indicates a flat uniform prior within the specified range, the nuisance parameter priors are listed in the bottom section of the table. Unless stated otherwise, the priors on the cosmological and clustering nuisance parameters match those of 
\citet{bosseboss_lcdm}.} 
	\label{tab:priors}
	\begin{tabular}{cc} 
		\hline
		Parameter & Prior \\
		\hline
		$\Omega_{\rm{b}}h^2$ & $U(0.019, 0.026)$ \\
		$\Omega_{\rm{c}}h^2$ & $U(0.01, 0.2)$ \\
		100$\theta_{\rm{MC}}$ & $U(0.5, 10.0)$ \\
		$\tau$ & $U(0.01, 0.8)$ \\
		$\rm{ln}( 10^{10}A_{\rm{s}})$ & $U(1.5, 4.0)$ \\
		$\rm{n}_s$ & $U(0.5, 1.5)$ \\
		$w$ & $U(-3 -0.3)$\\
		$w_{\rm{a}}$ & $U(-2, 2)$\\
		$\Omega_{\rm{K}}$ & $U(-0.3, 0.3)$\\
		$\sum m_{\nu}$ & $U(0.0, 2.0)$\\
		
		\hline
		$b_1$ & $U(0.5, 9.0)$\\
		$b_2$ & $U(-4, 8.0)$\\
		$a_{\rm{vir}}$ & $U(0.0, 12.0)$\\
		$\sigma_{\rm{err}}$ (eBOSS only) & $U(0.01, 6.0)$\\
		\hline

	\end{tabular}
\end{table}

\begin{table*}
	\centering
	\caption{Marginalised posterior constraints (mean values with 68 per-cent confidence interval, for $\sum m_{\nu}$ - 95 per-cent confidence interval) derived from Planck CMB and the full 
	shape analysis of BOSS + eBOSS clustering measurements on their own, as well as in combination with each other and with Pantheon supernovae Ia measurements (SN). All of the models considered here vary the dark energy equation of state parameter $w$. In addition to this, $w_a$CDM also varies $w_a$, allowing for the equation of state parameter that evolves with redshift, $wK$CDM varies curvature, and $w\nu$CDM varies neutrino mass sum $\sum m_{\nu}$. Note that for $wK$CDM the joint BOSS+eBOSS+Planck constraints should be interpreted bearing in mind that BOSS+eBOSS and Planck are discrepant in this parameter space (Figure \ref{fig:wklcdm}). }
	
	\label{tab:constraints}
	\begin{tabular}{c|c|cccc} 
		\hline
		\multicolumn{1}{c|}{} & \multicolumn{1}{c|}{} & \multicolumn{1}{c|}{Planck} & \multicolumn{1}{c|}{BOSS+eBOSS} &  \multicolumn{1}{c|}{BOSS+eBOSS+Planck} & \multicolumn{1}{c|}{BOSS+eBOSS+Planck+SN} \\
		\hline
		\multirow{3}{3em}{$w$CDM}   & $\sigma_{12}$ & $0.816\pm 0.011$ & $0.775^{+0.055}_{-0.066}$ & $0.804\pm 0.010$ & $0.8023\pm 0.0097$ \\
	                            	& $\omega_{\rm{DE}}$ & $0.509^{+0.15}_{-0.054}$ & $0.352^{+0.033}_{-0.044}$ & $0.341^{+0.020}_{-0.023}$ & $0.329\pm 0.012$\\
		& $w$ & $-1.41^{+0.11}_{-0.27}$ & $-1.10\pm 0.13 $ & $-1.066^{+0.057}_{-0.052}$ & $-1.033\pm 0.031$ \\
		\hline
		\multirow{4}{3em}{$w_a$CDM}   & $\sigma_{12}$ & $0.816\pm 0.012$ & $0.768^{+0.053}_{-0.061}$ & $0.807\pm 0.011$ & $0.805\pm 0.010$\\
	                            	&$\omega_{\rm{DE}}$ & $0.494^{+0.17}_{-0.062}$ & $0.356^{+0.042}_{-0.059}$ & $0.322^{+0.026}_{-0.039}$ & $0.330\pm 0.012$  \\
		& $w_0$ & $-1.22^{+0.33}_{-0.39}$ & $-1.09\pm 0.30$ & $-0.87^{+0.27}_{-0.22}$ & $-0.955\pm 0.086$  \\
			& $w_a$ & $< -0.330$ & $-0.13^{+1.1}_{-0.94}$ & $-0.60\pm 0.68$ & $-0.34^{+0.36}_{-0.30}$ \\
		\hline
		\multirow{4}{3em}{$wK$CDM}   & $\sigma_{12}$ & $0.896\pm 0.029$ & $0.754^{+0.056}_{-0.062}$ & $0.809\pm 0.011$ & $0.804\pm 0.010$\\
	                            	& $\omega_{\rm{DE}}$ & $0.323^{+0.073}_{-0.20}$ & $0.394^{+0.046}_{-0.053}$ & $0.346^{+0.020}_{-0.024}$ & $0.327\pm 0.012$ \\
		& $w$ & $-1.57^{+0.67}_{-0.38}$ & $-0.921^{+0.15}_{-0.093}$ & $-1.108^{+0.078}_{-0.067}$ & $-1.044\pm 0.036$ \\
			& $\omega_K$ & $-0.0116^{+0.0029}_{-0.0036}$ & $-0.057\pm 0.037 $ & $-0.0012\pm 0.0013$ & $-0.0006\pm 0.0011$\\
		\hline
		\multirow{4}{3em}{$w\nu$CDM}   & $\sigma_{12}$ & $0.810^{+0.019}_{-0.012}$ & $0.767^{+0.053}_{-0.064}$ & $0.796^{+0.016}_{-0.012}$ & $0.799^{+0.014}_{-0.011}$ \\
	                            	& $\omega_{\rm{DE}}$ & $0.508^{+0.15}_{-0.061}$ & $0.353^{+0.036}_{-0.046}$ & $0.346^{+0.020}_{-0.025}$ & $0.329\pm 0.012$ \\
		& $w$ & $-1.43^{+0.16}_{-0.26}$ & $-1.16^{+0.16}_{-0.13}$ & $-1.102^{+0.086}_{-0.058}$ & $-1.040^{+0.038}_{-0.033}$\\
			& $\sum m_{\nu}$ (eV) & $< 0.321$ & $< 1.34$ & $< 0.300$ & $< 0.211$ \\
		\hline
	\end{tabular}

\end{table*}

As in \cite{bosseboss_lcdm}, we are interested in constraining the cosmological parameters that are defined through physical units (i.e., avoiding $h$), as many standard cosmological parameter definitions imply averaging over the recovered posterior of $h$ \citep[for a more in-depth discussion see][]{littleh, bosseboss_lcdm}. This means that first, instead of describing the clustering amplitude today via $\sigma_8$, we use its equivalent defined on the physical scale of 12 Mpc, $\sigma_{12}$, as suggested by \cite{littleh}, and second, we use physical ($\omega_i$) rather the fractional ($\Omega_i$) densities of different components $i$ of the energy budget of the Universe with the two quantities related through:
\begin{equation}
\Omega_i = \omega_i/h^2.
\label{eq:Omegas}
\end{equation}
When considering extended cosmologies, we expect our chosen parameter space to be most relevant for the cases where the $\Lambda$CDM assumptions about dark energy are relaxed, as physical dark energy density $\omega_{\rm{DE}}$ is not well constrained by the CMB or large scale structure probes and depends on the assumed dark energy model. This statement is especially true for Planck, which probes the Universe at the redshift of recombination. The dimensionless Hubble parameter, $h$, is defined by the sum of all energy contributions from baryons ($\omega_{\rm b}$), cold dark matter ($\omega_{\rm c}$), neutrinos ($\omega_{\nu}$), dark energy, and curvature ($\omega_{K}$) : 
\begin{equation}
h^2 = \omega_{\rm b} + \omega_{\rm c} + \omega_{\nu} + \omega_{\rm DE} + \omega_{K},
\label{eq:hubble}
\end{equation}
with dark energy comprising the majority of the total energy budget today. Therefore, when we introduce additional freedom to dark energy modelling, this is also reflected in the constraints on $h$ and any parameter that is defined through it. In this analysis we, therefore, allow the dark energy equation of state parameter $w=p_{\rm{DE}}/\rho_{\rm{DE}}$ deviate from its $\Lambda$CDM value of $w=-1$ and treat it as a free parameter for all extensions considered in order to explore the effects on physical parameter space constraints. 

In addition to the basic $w$CDM model with constant $w$, we also consider a more general parametrization where $w$ is allowed to evolve with the scale factor $a$ \citep{Chevallier, Linder}:
\begin{equation}
w = w_0+w_a(1-a).    
\label{eq:wa}
\end{equation}
Here $w_0$ and $w_a$ are free parameters; we refer to this case as $w_{a}$CDM model. We also explore $wK$CDM - non-flat models with $\Omega_k\neq0$. Here, as with the other energy budget components, we are interested in physical curvature density $\omega_{k}=\Omega_kh^2$. Finally, we investigate the constraints on neutrino mass sum $\sum m_{\nu}$, by allowing it to vary freely instead of fixing it to the fiducial value of $\sum m_{\nu}=0.06$eV, corresponding to the minimum value allowed by neutrino oscillation experiments under normal hierarchy \citep{mnu}. We refer to this model as $w\nu$CDM.

We use \textsc{CosmoMC} \citep{Lewis_2002} to perform Monte Carlo Markov chain (MCMC) sampling. For the linear-theory matter power spectrum prediction, \textsc{CosmoMC} uses CAMB \citep{Lewis_2000}, adapted to compute the theoretical model for anisotropic clustering measurements, as described in Sec. \ref{sec:model}. We sample over the basis parameters used by \textsc{CosmoMC}:
\begin{equation}
\pmb{\theta}_{\rm base} = \left(\omega_{\rm b},\omega_{\rm c},\Theta_{\rm MC}, A_{\rm s}, n_{\rm s}, w_0, w_a, \Omega_k, \sum m_{\nu} \right),
\label{eq:base_cosmomc}
\end{equation}
where $\Theta_{\rm MC}$ is 100 times the approximate angular
size of the sound horizon at recombination. For each of the models described in this section we only vary the relevant extended parameters, fixing the rest to their fiducial values, as described above. We impose flat and uninformative priors, except for $\omega_{\rm{b}}$, with the priors for $\Lambda$CDM parameters matching those of \cite{bosseboss_lcdm}. Our flat prior for $\omega_{\rm{b}}$ informs clustering measurements, as they do not constrain this parameter on their own, and is 25 times wider than the corresponding Planck constraint. We also need to specify the allowed values of the Hubble parameter, $h$. In order to be consistent with \cite{bosseboss_lcdm}, we choose the same range of $0.5<h<0.9$. While this range is somewhat restrictive for Planck on its own for varying dark energy cosmologies, these limits have little effect on the physical parameter space that we consider, and are mostly uninformative once Planck is combined with clustering or when considering clustering alone. Finally, these limits are motivated by the direct measurements of $H_0$, which fall well within this range \citep{Riess2021, cosmicflows, tdcosmo, megamasers, frb}. A summary of all cosmological priors used in this analysis is presented in Table \ref{tab:priors}.

\section{Results}
\label{section:results}

We are mainly interested in clustering constraints from BOSS+eBOSS as well as their combination with Planck. Where found informative, we supplement the clustering constraints with those from SNe Ia. The summary of our results on the main parameters of interest is shown in Table \ref{tab:constraints} with further constraints available in Appendix \ref{appx:A}.

\subsection{Evolving dark energy - wCDM}
\label{section:wcdm}

\begin{figure*}
\includegraphics[width=0.61\textwidth]{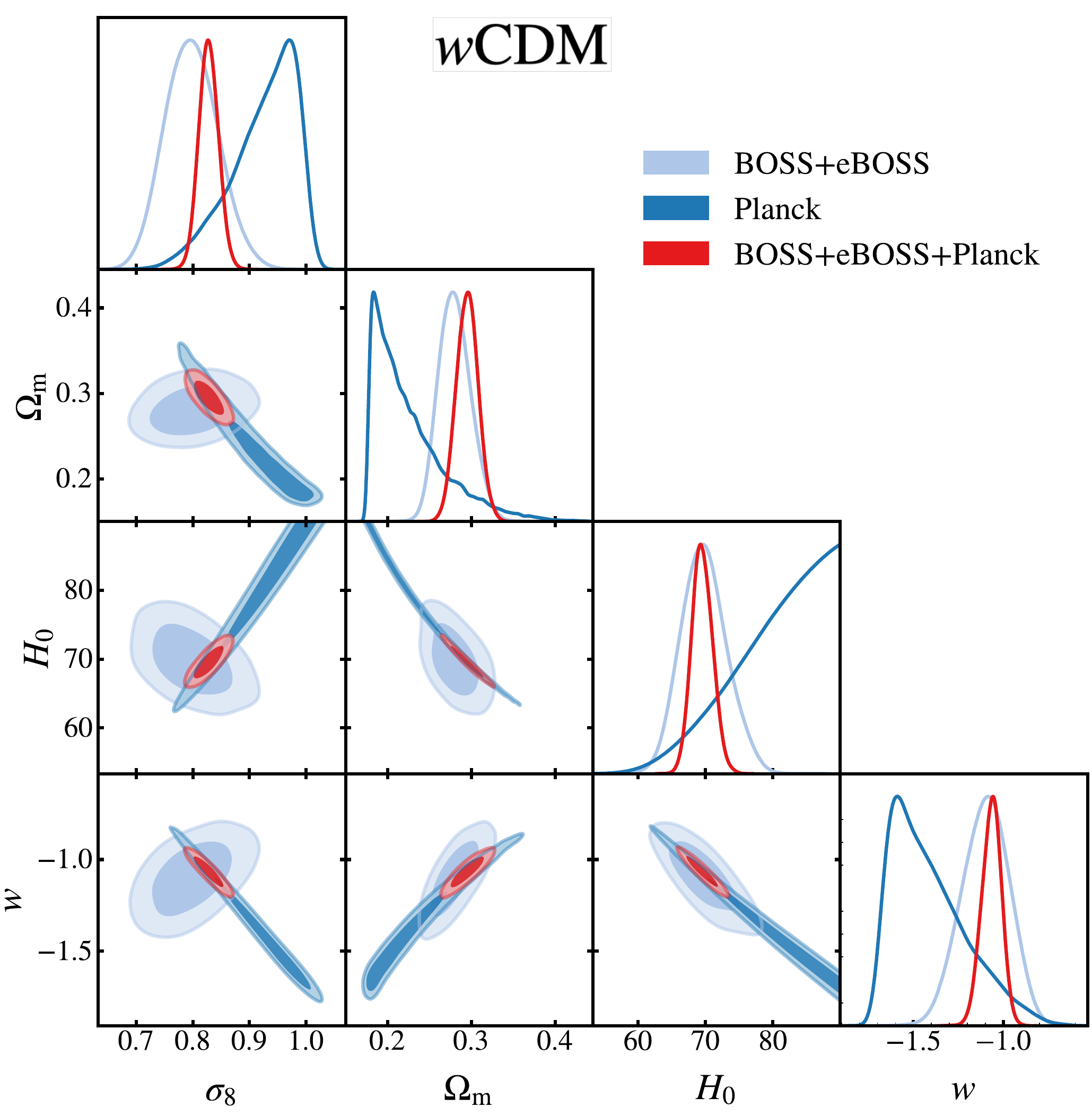}
\includegraphics[width=0.61\textwidth]{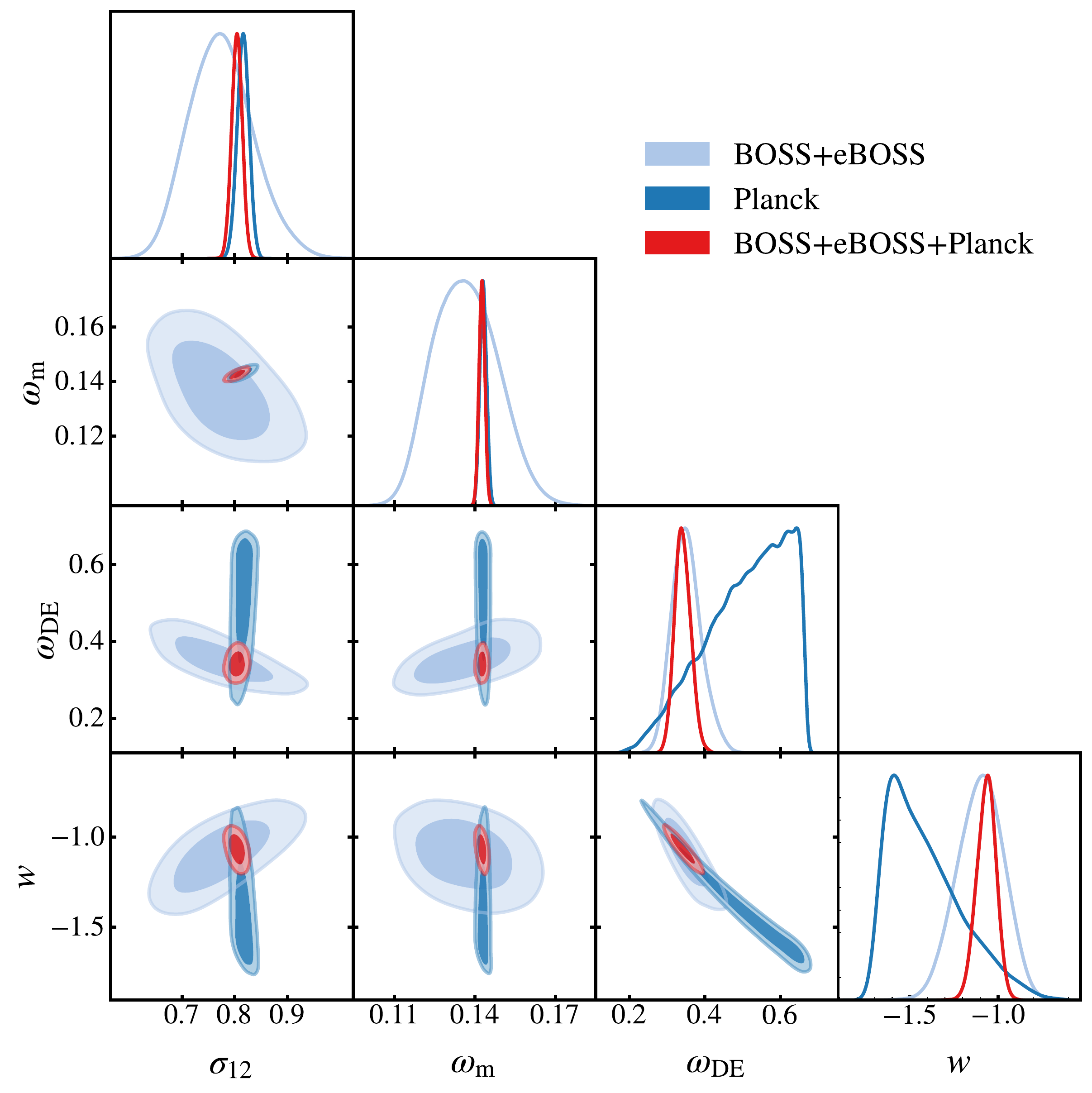}
\caption{Marginalised posterior contours in the `traditional' and $h$-independent parameter spaces 
derived from the full shape of anisotropic clustering measurements of BOSS DR12 galaxies in 
combination with eBOSS quasars (light blue) and CMB measurements by Planck (dark blue) for 
a $w$CDM model. The joint constraints are shown in red. In physical parameter 
space, Planck is able to constrain the clustering amplitude today $\sigma_{12}$ even 
in models with free $w$.}
\label{fig:wcdm}
\end{figure*}

\begin{figure}
\centering
\includegraphics[width=0.95\columnwidth]{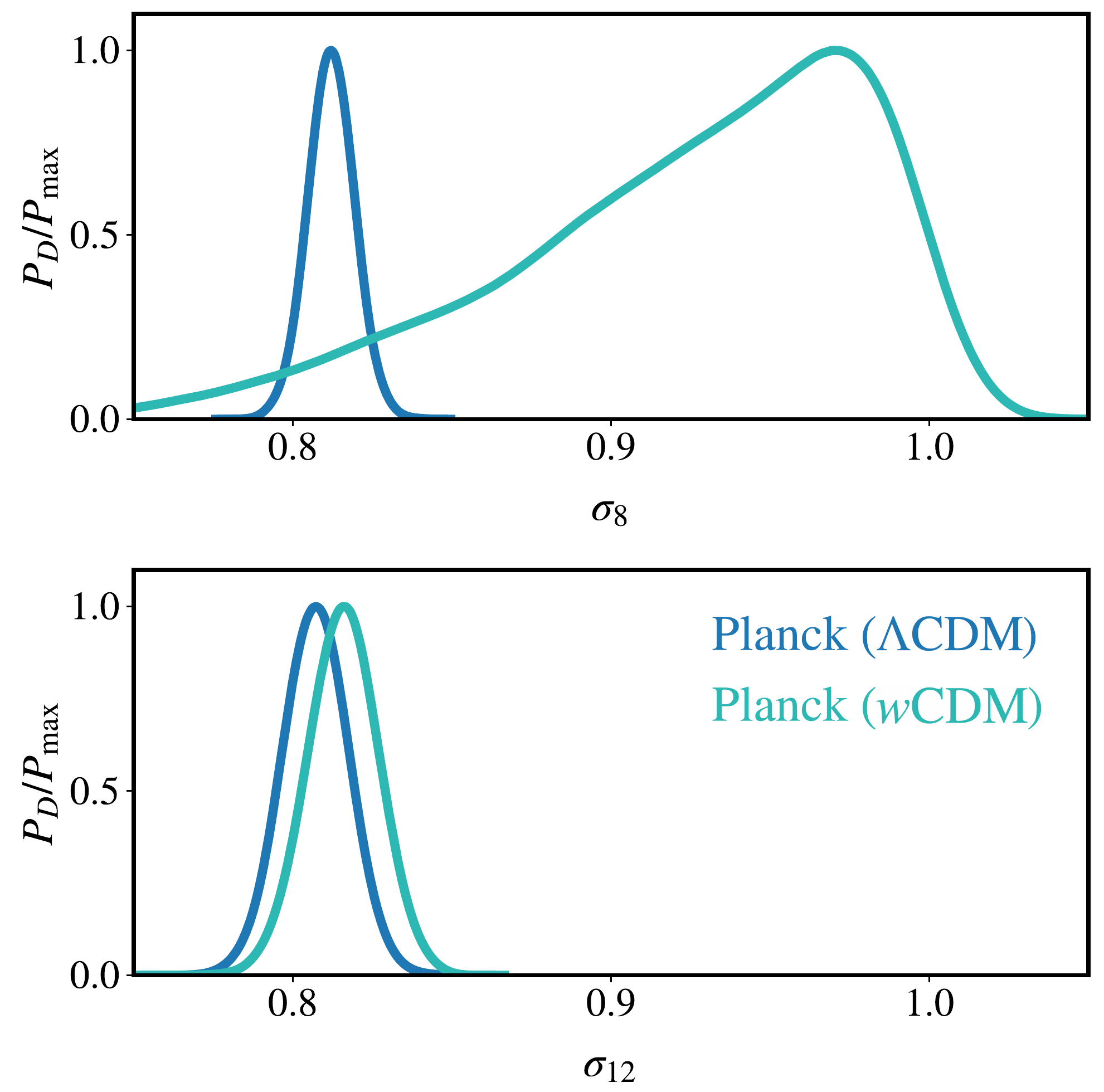}
\caption{ Upper panel: one-dimensional marginalised Planck posteriors for $\sigma_8$ for $\Lambda$CDM (blue) and $w$CDM (green) cosmologies . Lower panel: the corresponding posteriors of $\sigma_{12}$. The difference between two panels is due to $h^{-1}$Mpc units used to define the scale at which the linear density field variance is measured for $\sigma_8$. As Planck does not constrain $H_0$ well once $w$ is allowed to freely vary, the resulting posterior, over which the clustering amplitude is effectively averaged, is extremely wide and results in degraded constraint on $\sigma_8$. The parameter $\sigma_{12}$ is not affected by this issue, as it is defined on a Mpc scale - the precision of the measurement for Planck is the same in both cosmologies shown. }
\label{fig:s8vss12}
\end{figure}

$\Lambda$CDM assumes a cosmological constant-like behaviour for dark energy with a fixed $w=-1$. Nevertheless, $w$ may be allowed to deviate from this value and be treated as a free parameter. As the simplest of the models considered here, we will use our results for $w$CDM to illustrate the behaviour of our data in the physical parameter space for this class of cosmologies. Figure \ref{fig:wcdm} presents our constraints from BOSS+eBOSS and Planck on their own (light and dark blue respectively) as well as their combination (in red) with constraints on the standard parameter space shown in the upper panel for comparison.

Comparing the two sets of panels in Figure \ref{fig:wcdm}, it is clear that the posterior degeneracy directions for $H_0$ (whose value is determined by the sum of the physical densities of all the components) are set by the $\omega_{\rm{DE}}$ component. For Planck, the constraint on $\omega_{\rm{DE}}$ is set by the prior - as discussed before, Planck does not probe the redshfits at which dark energy becomes dominant directly, but rather is able to provide model-dependent constraints based on $D_{\rm A}(z_{\star})$. The CMB observations, therefore, do not constrain dark energy density once the evolution of this component is not well defined.

In contrast, $\omega_{\rm{m}}$ is set by the scale dependence on the amplitude of CMB spectra (for a fixed acoustic angular scale) and is not sensitive to the assumptions on dark energy. Following $H_0$, any parameter defined through it also exhibits similar degeneracies, as they are effectively averaged over the posterior of $H_0$ - as a result, $\sigma_8$ and $\Omega_{\rm{m}}$ are not well constrained by Planck either.

Nonetheless, importantly, that does not mean that Planck is unable to measure the clustering amplitude today - the $w$CDM constraint on $\sigma_{12}$ has the same precision as in $\Lambda$CDM model, as illustrated in Figure \ref{fig:s8vss12}. The lack in constraining power on $\sigma_8$ is, therefore, an artefact of using $h^{-1}$Mpc units. To understand why Planck does not lose constraining power on $\sigma_{12}$ even in extended cosmologies, one can first note that $w$ and $\sigma_{12}$ show almost no correlation for this probe. This behaviour arises because the change in $w$ is compensated by a change in $\omega_{\rm{DE}}$, as is evident from Planck's constraints in the $w-\omega_{\rm{DE}}$ plane: only certain combinations of these parameters, set by $D_{\rm A}(z_{\star})$, are allowed, with the resulting degeneracy corresponding to a constant $\sigma_{12}$. This result means that a preference for more negative $w$ closely corresponds to an increase in $\omega_{\rm{DE}}$.

Our analysis is, therefore, the first one to quote a CMB constraint on clustering amplitude today in cosmologies with varying $w$. The Planck best fit value of $\sigma_{12}=0.816 \pm 0.011$ that we find for $w$CDM model is slightly higher than that for $\Lambda$CDM ($0.807\pm 0.011$) - this increase is because the higher values of $\omega_{\rm DE}$ allowed in $w$CDM correspond to a more negative $w$, thus the dark energy content is lower at the start of the epoch when this component becomes relevant, which results in slightly more total structure growth. Nevertheless, as $\sigma_{12}$ is mostly determined by the physical matter density, as discussed above, this shift in the clustering amplitude value is minimal.

The advantage of our physical parameter space is most evident for Planck due to its lack of constraining power on $H_0$; nonetheless, even for clustering probes, the precision on physical parameter constraints degrades less, compared to their $h$-dependent counterparts, once dark energy model assumptions are relaxed. 

While clustering on its own prefers a mean equation of state parameter value that is compatible with $w= -1$ ($w=-1.10 \pm 0.13$), for Planck and the combination BOSS+eBOSS+Planck, the fiducial value is just outside of the 68\% confidence limit. This result appears because of the significant volume of Planck's posterior corresponding to models with high dark energy content, which shifts the mean $w$. The addition of low redshift information from clustering rules out such models and brings the joint constraints closer to $\Lambda$CDM. 

Comparing our BOSS+eBOSS+Planck constraints with previously published full shape analysis of BOSS clustering wedges by \cite{sanchez17}, who obtain $w=-0.991^{+0.062}_{-0.047}$ for BOSS+Planck (2015), reveals that our updated analysis shifts the mean $w$ by $\sim 1\sigma$  towards more negative values. This result may be attributable to a number of differences between the analyses, most notably the updated CMB measurements from Planck. 

In addition to this approach, \cite{Brieden2022} performed a reconstructed power spectrum multipole analysis of BOSS DR12 LRG and eBOSS QSO samples. In this analysis, the information from BAO and RSD summary statistics is complemented by additional summary statistic derived from the shape of the power spectrum \citep[ShapeFit, ][]{shapefit}. Our BOSS+eBOSS(+Planck) constraint $w=-1.10\pm0.13$ ($w=-1.066^{+0.057}_{-0.052}$) agrees well with that of \cite{Brieden2022}: $w=-0.998^{+0.085}_{-0.073}$ ($w=-1.093^{+0.048}_{-0.044}$), showing the robustness of these results.

\subsection{Evolving dark energy equation of state - $w_{a}$CDM}

\begin{figure}
\includegraphics[width=0.99\columnwidth]{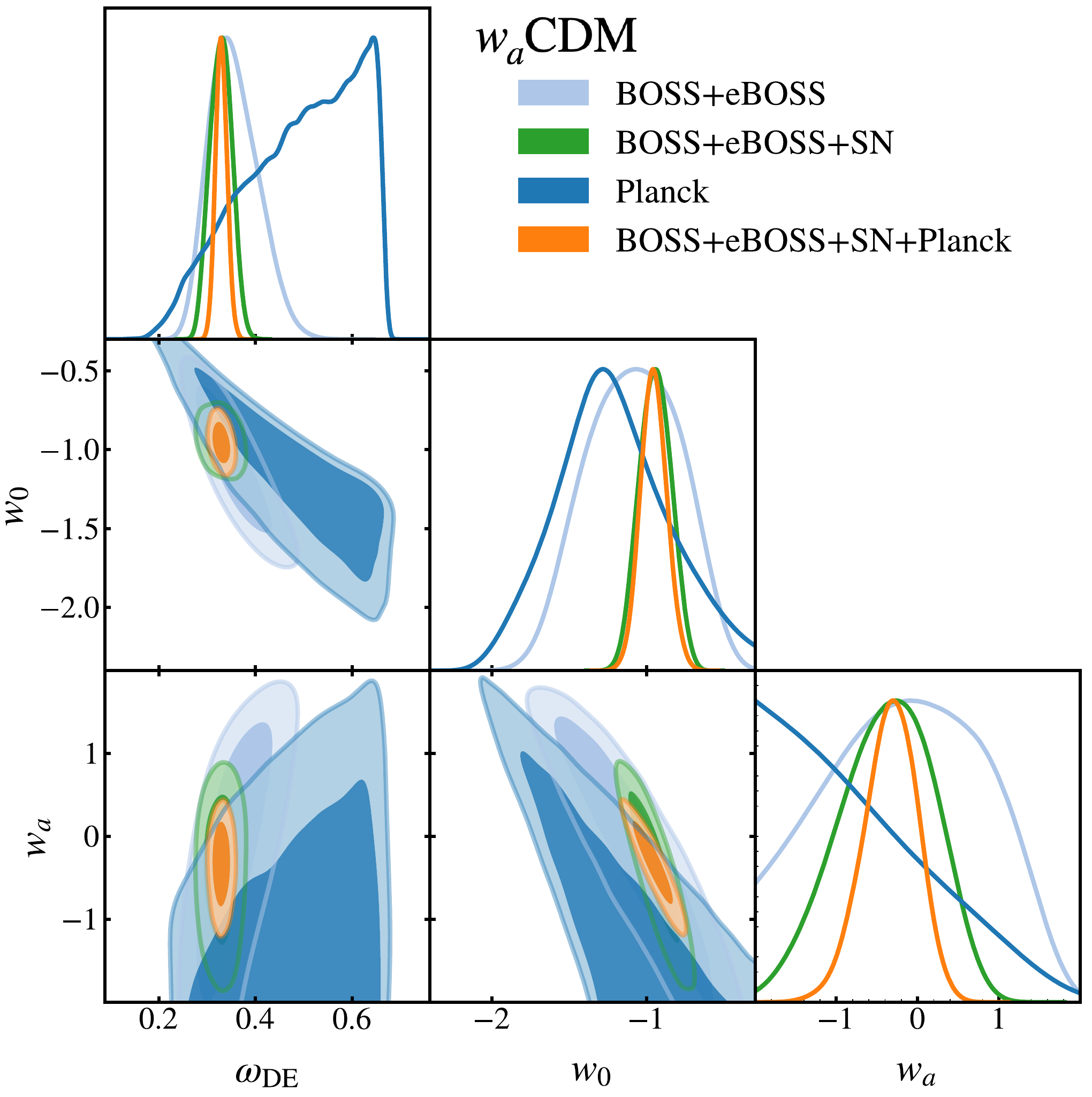}
\caption{Marginalised posterior contours for dark energy parameters in $wa$CDM, where the dark energy equation of state parameter $w$ is allowed to evolve in time, as defined in equation~(\ref{eq:wa}). We show constraints from the full-shape clustering analysis of BOSS DR12 galaxies in combination with 
eBOSS quasars (light blue), their combination with Pantheon SN Ia measurements (green), CMB 
constraints by Planck (in dark blue), and the combination of all four datasets (in orange).  }
\label{fig:wacdm}
\end{figure}

\begin{figure*}
\includegraphics[width=0.7\textwidth]{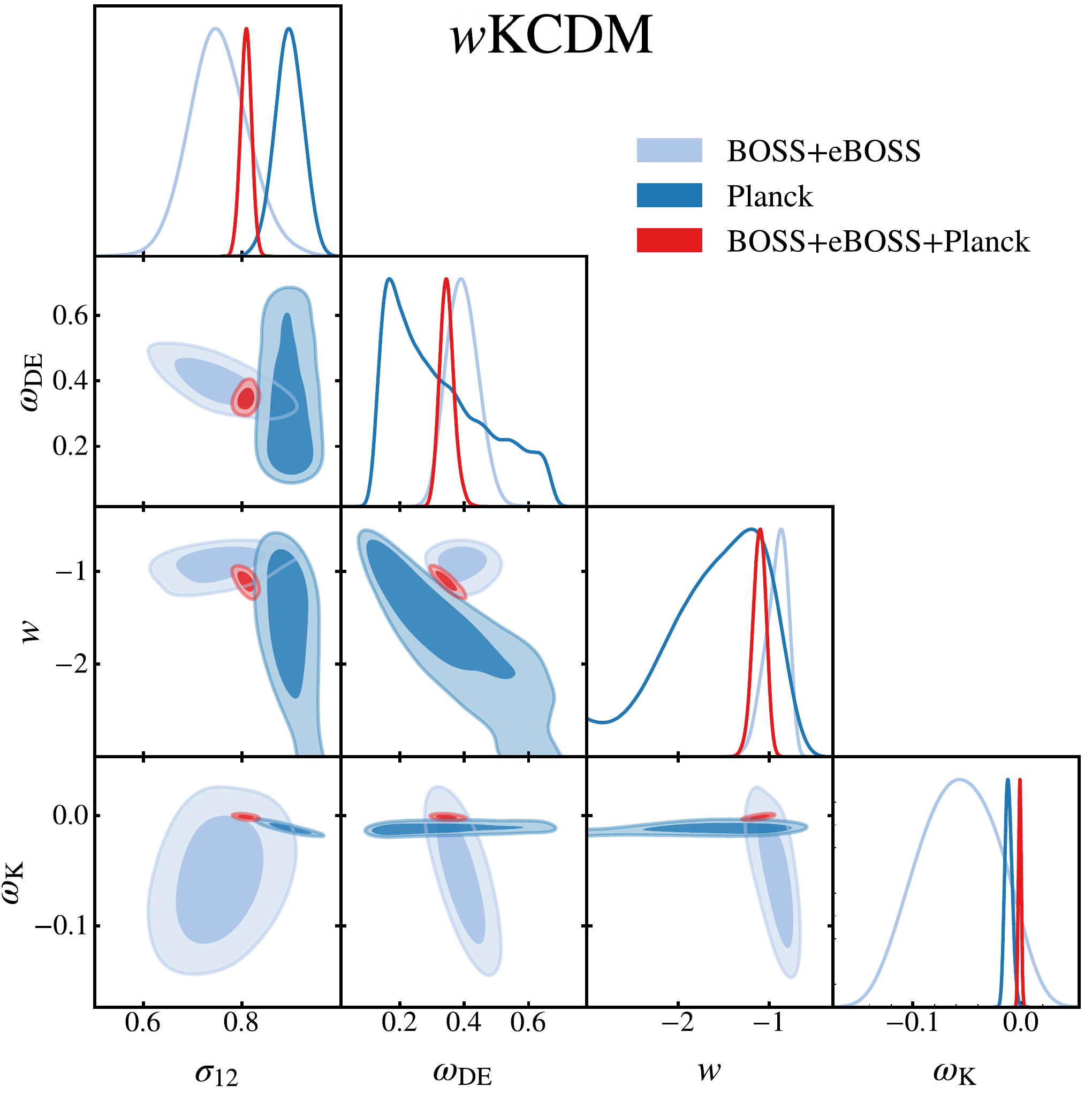}
\caption{Marginalised posterior contours for dark energy parameters and physical matter and curvature densities in $wK$CDM, where, in addition to varying $w$, we also allow for non-zero curvature. The constraints from the full-shape clustering analysis of BOSS DR12 galaxies in combination with eBOSS quasars are shown in light blue, CMB constraints by Planck are displayed in dark blue and the combination of the two sets of probes is shown in red. Note the discrepancy between Planck and BOSS+eBOSS in $\sigma_{12}$ as well as $w$-$\omega_{\rm{DE}}$ plane. }
\label{fig:wklcdm}
\end{figure*}

We further generalise the dark energy description allowing its equation of state parameter to vary with time, as defined in equation (\ref{eq:wa}). Figure~\ref{fig:wacdm} shows our constraints for the dark energy parameters: $w, w_{\rm a}, \omega_{\rm DE}$. Here, we combine our clustering constraints with SNe Ia, which provide background constraints for the lowest redshifts and are, therefore, extremely useful for probing the evolution of dark energy. 

The additional freedom in the equation of state model has minimal impact on the constraints on $\sigma_{12}$ and $\omega_{\rm{DE}}$. However, the addition of SNe Ia data halves the error on dark energy density with the resulting constraint of $\omega_{\rm{DE}}=0.330 \pm 0.012$. All of the dataset combinations considered recover a value of $w_0$ that is consistent with -1, although with significantly larger uncertainty than in $w$CDM. Planck does not constrain $w_{\rm{a}}$ on its own, but combining it with the clustering and supernovae data yields a value compatible with no evolution (for BOSS+eBOSS+Planck+SN $w_a=-0.34^{+0.36}_{-0.30}$). 

We may compare our constraints with those from the completed SDSS consensus analysis by \cite{eboss2021}, although note that they use additional datasets, including eBOSS luminous red galaxy and emission line galaxy samples as well as BAO from the Ly$\alpha$ forest, and use summary statistics for BAO and RSD to obtain the joint constraints. The consensus analysis also uses reconstructed BAO, while we perform no reconstruction in this work. The quoted constraints for combined  Planck+Pantheon SNe, SDSS BAO+RSD and DES $3~\times~2$pt data are $w_0=-0.939\pm0.073$ and $w_{\rm{a}}=-0.31^{+0.28}_{-0.24}$, which are in excellent agreement with our BOSS+eBOSS+Planck+SN result. 

\cite{Chudaykin2021} also performed a full shape analysis using a model based on the Effective Field Theory of Large Scale Structure \citep[EFT,][]{eft}. They analysed BOSS DR12 luminous red galaxy redshift space power spectrum multipoles in combination with BAO measurements from post-reconstructed power spectra of BOSS DR12 supplemented with a number of additional BAO measurements from SDSS, including those from eBOSS QSO sample and additionally augmented by adding supernovae Ia measurements from Pantheon (as also done in this work). Their final constraints of $w_0=-0.98^{+0.10}_{-0.11}$ and $w_{\rm{a}}=-0.32^{+0.63}_{-0.48}$ are tighter than ours (most likely due to the additional BAO data), but in an excellent agreement with our BOSS+eBOSS+SN results: $w_0=-0.94^{+0.20}_{-0.19}$ and $w_{\rm{a}}=-0.40^{+1.0}_{-1.2}$.

\subsection{Non-zero curvature - $wK$CDM}

We explore what occurs when, in addition to varying $w$ (but with $w_{\rm{a}}=0$), we also allow for non-flat models. The resulting constraints are shown in Figure \ref{fig:wklcdm}. Here, together with the dark energy parameter constraints, we also display the physical curvature density $\omega_{\rm{K}}=\Omega_{\rm{K}}h^2$. 

It is interesting that the Planck data constrain the physical curvature well, with the mean value of  $\omega_{\rm{K}}=-0.0116^{+0.0029}_{-0.0036}$, indicating a strong preference for non-zero curvature. We compare this result with the constraint on the $h$-dependent equivalent, $\Omega_{\rm{K}}=-0.030^{+0.018}_{-0.010}$; note how physical units allow us to detect the deviation from flatness at a higher significance (4$\sigma$ for $\omega_{\rm{K}}$ versus 1.6$\sigma$ for $\Omega_{\rm K}$). This preference for closed Universe is a known feature of Planck data and believed to be related to the lensing anomaly \citep{Planck2018}. Nonetheless, our physical curvature constraint indicates the most significant deviation from flat Universe yet, which is especially interesting bearing in mind that, in addition to curvature, we are also varying $w$ and would, therefore, expect a somewhat more significant preference for closed Universe in fixed $w=-1$ case (however, as seen in Figure \ref{fig:wklcdm}, dark energy parameters are almost independent of $\omega_{\rm K}$ for Planck; therefore, we expect the change in the result to be minimal).

Recently, \cite{clusterK} reported that clustering data alone may show a $2\sigma$ preference for a closed universe. There, a full shape analysis based on EFT is performed on the power spectra multipoles from the full combined 6dFGS, BOSS, and eBOSS catalogues. The analysis by \cite{Chudaykin2021}, however, finds a less significant deviation of $\sim1\sigma$. Neither of these analyses vary $w$, which allows for somewhat tighter constraints than what we expect for $wK$CDM (as seen in Figure \ref{fig:wklcdm}, clustering exhibits some degeneracy between the two parameters). The mean value of $\omega_{\rm{K}}$ preferred by BOSS+eBOSS in our analysis, $\omega_{\rm{K}}=-0.057\pm 0.037$, also deviates from 0, but is consistent with flatness at 95\% confidence level, indicating no significant preference for a closed universe. In terms of dark energy constraints, the effect of allowing a free varied curvature for clustering is to allow for larger values of $\omega_{\rm{DE}}$.

CMB and clustering datasets are highly complementary in $wK$CDM, with Planck providing a measurement on curvature and BOSS+eBOSS constraining dark energy parameters. Nevertheless, the two datasets are discrepant within this cosmology. This behaviour is most clear from the $w-\omega_{\rm{DE}}$ projection in which the $2\sigma$ regions of the two sets of contours show little overlap. As discussed before, this particular degeneracy is defined by the clustering amplitude today, so this discrepancy is also reflected in the $\sigma_{12}$ constraints, with Planck preferring a $2.4\sigma$ higher value. This model displays the greatest shift in Planck's $\sigma_{12}$ out of all of the cosmologies considered in this work.

\begin{figure}
\includegraphics[width=0.99\columnwidth]{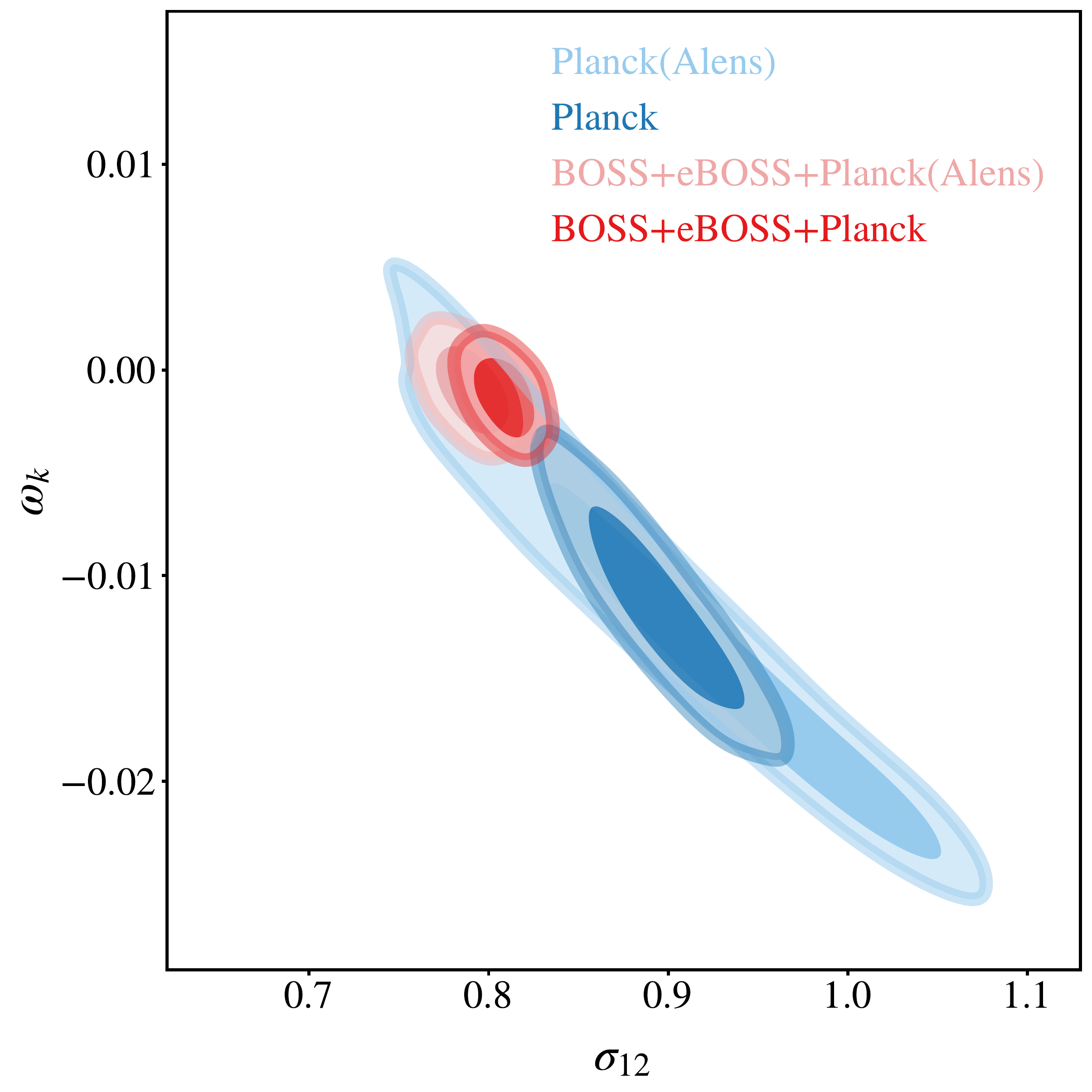}
\caption{Marginalised posterior contours for $\sigma_{12}$ and $\omega_{\rm K}$ from Planck and its combination with BOSS+eBOSS in $wK$CDM. We compare two cases - one with varying $A_{\rm{lens}}$ (light blue and pink) and a fiducial one, with $A_{\rm{lens}}$ fixed to 1 (dark blue and red).}
\label{fig:wklcdm_Alens}
\end{figure}

A discrepancy between Planck and BAO measurements and Planck and full shape analysis of clustering measurements was also found in previous work that varied curvature but kept $w$ fixed \citep{valentino2020, Handley2021, Vagnozzi2021, clusterK}. Our analysis, therefore, demonstrates that a degradation of constraining power when varying $w$ does not provide a solution to this tension. 

Planck's lensing anomaly, which is related to preference for non-zero curvature, can be characterised 
by the phenomenological parameter $A_{\rm{lens}}$, which scales the amplitude of the lensing power relative to the physical value. In the absence of systematics or non-standard physics, 
$A_{\rm{lens}}=1$ and is highly degenerate with the measured cosmological parameters that set the amplitude of the power spectrum at late times - $\sigma_{12}$ and $\omega_{\rm K}$. Figure~\ref{fig:wklcdm_Alens}, illustrates how allowing $A_{\rm{lens}}$ to freely vary extends Planck's posterior contours to include $\omega_{\rm K}=0$ within $2\sigma$ and, therefore, recovers flat $\Lambda$CDM. The extension of Planck's posterior distribution of $\sigma_{12}$ to lower values allows for a reconciliation with the constraints from BOSS+eBOSS and reduces the mean $\sigma_{12}$ inferred from the combined BOSS+eBOSS+Planck analysis. Varying $A_{\rm{lens}}$ allows to compensate for the excess lensing and brings Planck in line with clustering measurements in $wK$CDM. This behaviour is expected and is consistent with existing analyses (for example, \cite{diValentino21} demonstrated how varying $A_{\rm{lens}}$ allows Planck to be more consistent with flatness and brings it to a better agreement with the BAO measurements for cosmological models with varying curvature, $w$, and neutrino mass sum). Here we additionally note that the inclusion of $A_{\rm{lens}}$ does significantly degrade Planck's ability to constrain $\sigma_{12}$.

\subsection{Massive neutrinos - $w\nu$CDM}

\begin{figure}
\includegraphics[width=0.99\columnwidth]{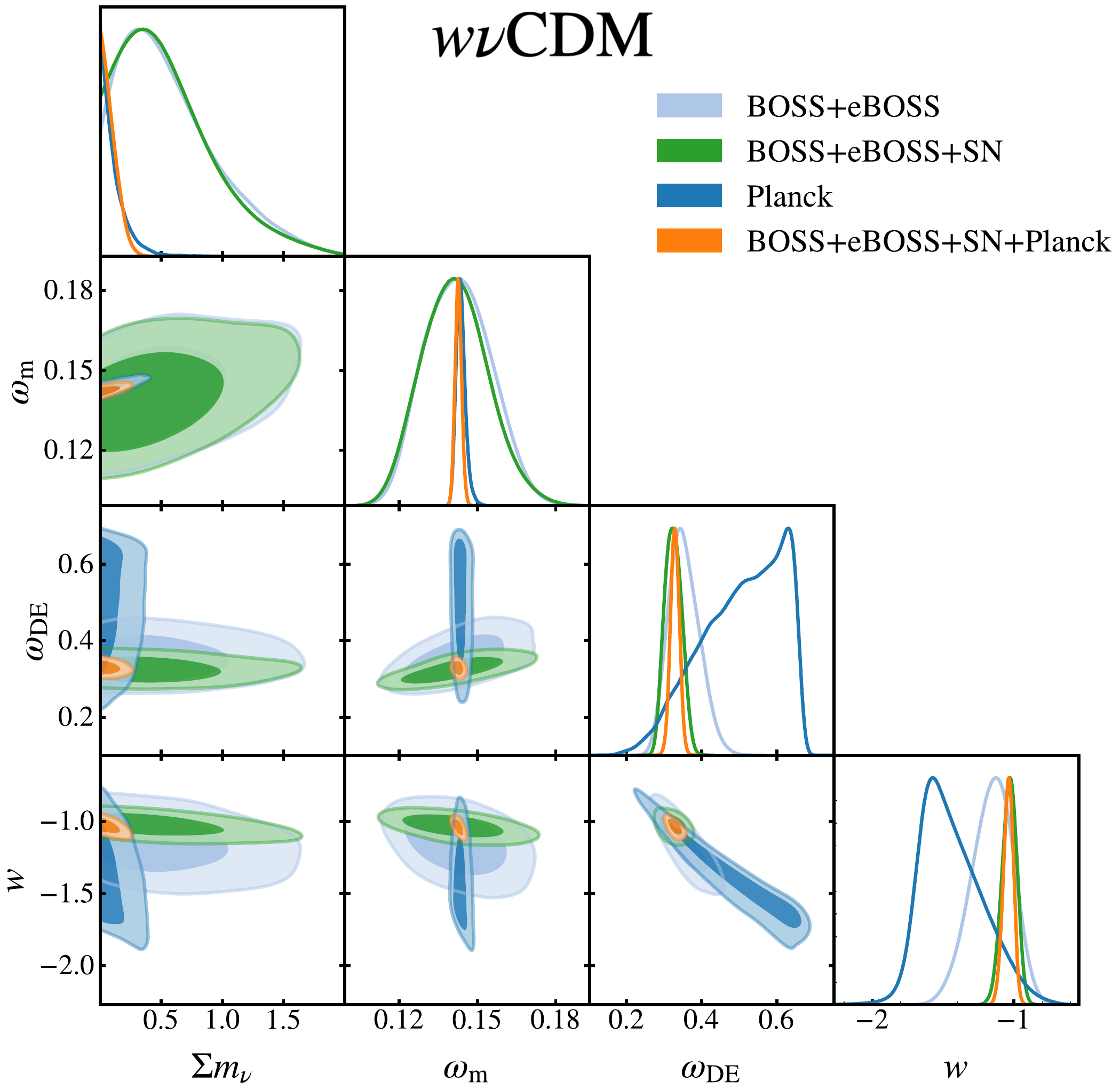}
\caption{Marginalised posterior contours for dark energy parameters and the neutrino mass sum $\sum m_{\nu}$ in $w\nu$CDM. Here, we explore varying $w$ together with $\sum m_{\nu}$. We show constraints from the full-shape clustering analysis of BOSS DR12 galaxies in combination with eBOSS quasars (light blue), their combination with Pantheon SN Ia measurements (green), CMB constraints by Planck (in dark blue) and the combination of all four datasets (in orange). }
\label{fig:wnucdm}
\end{figure}

Finally, we vary the neutrino mass sum, $\sum m_{\nu}$: here, once again, we supplement our clustering and CMB datasets with Pantheon supernovae. As $\sum m_{\nu}$ exhibits a degeneracy with the dark energy equation of state \citep[for a more detailed discussion see][]{Hannestad2005}, varying $w$ is expected to degrade the resulting constraints. The addition of SNe Ia, therefore, allows to improve the precision of our constraint through providing a measurement of $w$. 

While clustering alone does not provide a tight upper limit for the neutrino mass ($\sum m_{\nu} < 1.34$ eV from BOSS+eBOSS at 95\% confidence), its combination with Planck does allow for some improvement as compared with Planck alone ($\sum m_{\nu} < 0.300$ eV for joint constraints versus $\sum m_{\nu} < 0.321$ eV for Planck). It might, nevertheless, initially seem surprising that the improvement is rather minimal \citep[especially, compared to the constraints obtained from combined Planck full shape clustering analyses in $\Lambda$CDM, such as][]{Ivanov_mnu, Tanseri_mnu}. This result can be understood by noting that, due to the degeneracy between the physical matter density, $\omega_{\rm{m}}$, and $\sum m_{\nu}$, a precise measurement of $\omega_{\rm{m}}$ is required to shrink the upper limit on $\sum m_{\nu}$. While clustering constraints on either of these parameters are much looser than those of Planck, it may, nonetheless, be able to improve on Planck's measurement of $\omega_{\rm{m}}$ by excluding some of the cosmologies in  $\omega_{\rm{m}}-\omega_{\rm{DE}}$ space allowed by Planck. This effect happens to be more significant in a $\Lambda$CDM cosmology, where the full shape BOSS+eBOSS analysis yields a lower mean value of $\omega_{\rm m}$, as compared to Planck, resulting in a lower $\omega_{\rm m}$ from a combined measurement and, therefore, reducing the maximum limit of $\sum m_{\nu}$. As shown in Figure \ref{fig:wnucdm}, in the cosmology with varying $w$ the BOSS+eBOSS contour is perpendicular to Planck in $\omega_{\rm{m}}-\omega_{\rm{DE}}$, resulting in minimal effect on the constraint on $\omega_{\rm{m}}$ and, consequently, little improvement on the upper mass limit for $\sum m_{\nu}$.

Our tightest constraint then arises from the combination of BOSS+eBOSS+Planck+SN, which places the upper limit of $\sum m_{\nu}<0.211$ eV at 95\% confidence. This improvement is due to the fact that the supernovae constraint on $\omega_{\rm{DE}}$ does result in a tighter constraint on $\omega_{\rm{m}}$, which, in turn, shrinks the $\omega_{\rm{m}}$, and $\sum m_{\nu}$ degeneracy. 

The SDSS consensus analysis \citep{eboss2021} provides a lower neutrino mass sum limit for w$\nu$CDM of $\sum m_{\nu}<0.139$ eV (Planck+BAO+RSD+SN+DES, 95\% upper limit); nevertheless, these constraints do include additional data  (reconstructed BAO and RSD measurements from full SDSS data as well as measurements from the Dark Energy Survey, DES) and are, therefore, not directly comparable. As described above, the final limit is extremely sensitive to $\omega_{\rm{m}}$ and, therefore, not only the statistical power of a particular combination of measurements, but also the position and orientation of the contours (which may differ for full shape and BAO-only analyses).

\section{Conclusions}
\label{section:conclusions}

In this work we performed a full shape analysis of the anisotropic two-point clustering measurements from BOSS galaxy and eBOSS QSO samples together with Planck CMB and Pantheon SNe Ia measurements and explored extensions to the $\Lambda$CDM cosmological model. In particular, we were interested in models with free dark energy equation of state parameter $w$ and the resulting constraints in physical parameter space. 

We demonstrated that CMB  recovers a tight degeneracy in the $w-\omega_{\rm DE}$ parameter space and is able to constrain the linear density field variance, $\sigma_{12}$, as well as the physical curvature density $\omega_{\rm K}$, to high precision. The apparent lack of constraining power when using $\sigma_8$ is, therefore, only an artifact of using $h^{-1}$Mpc units when defining the scale on which the density field variance is measured. This approach results in averaging over the posterior of $h$ that Planck does not constrain in evolving dark energy models. We subsequently presented the first CMB measurements of the clustering amplitude today in cosmologies with varying $w$ and found that the clustering amplitude tends to increase for such models. This behaviour is because a more negative $w$ requires a lower initial $\omega_{\rm DE}$ value to reach the same constraint at redshift zero. This observation also rules out the evolving dark energy models considered here as a potential way to bring Planck's predicted amount of structure growth closer to weak lensing observations on their own, as such extensions do not affect the initial amplitude of matter fluctuations and the clustering amplitude today $\sigma_{12}$ is set by the $w-\omega_{\rm DE}$ degeneracy and is well measured even in the extended models. 

When, in addition to $w$, the curvature is also allowed to vary, BOSS+eBOSS and Planck become discrepant, most significantly in the $w-\omega_{\rm DE}$ plane and, subsequently, in the resulting values of $\sigma_{12}$, with Planck preferring a 2.4$\sigma$ higher value than BOSS+eBOSS. Varying dark energy models are, therefore, not able to bring the two probes in a better agreement for curved cosmologies. In addition to this result, our physical curvature density constraint for Planck $\omega_{\rm{K}}=-0.0116^{+0.0029}_{-0.0036}$ prefers a curved Universe at $4\sigma$ significance, which is $2.4\sigma$ higher than what is found using $\Omega_{\rm{K}}$. 

It is encouraging that the extended model constraints that we derive from our full shape analysis are compatible with a $\Lambda$CDM cosmology (with the greatest deviation seen for $\omega_{\rm K}$ but still within $\omega_{\rm K}=0$ at 95\% confidence) as well as with previous clustering analyses. We derive the 95\% upper limit for the neutrino mass sum of $\sum m_{\nu}<0.211$ eV (BOSS+eBOSS+Planck+SN), which is a higher value than that of the SDSS consensus analysis (though the two constraints are not directly comparable, as the consensus analysis makes use of a more extensive dataset). When $w$ is varied freely, clustering alone only allows for a modest improvement in the upper limit of $\sum m_{\nu}$.

Our analysis demonstrates the strength of the physical parameter space in constraining extended cosmologies. While we are currently still unable to place tight constraints on the dark energy parameters directly, we were able to show how even the high-redshift observations place limits on the allowed behaviours. We were also able to provide a consistent picture of the current state of full-shape clustering constraints, which were shown to be highly complimentary to the CMB measurements. With CMB providing information on physical matter and curvature densities, as well as setting a strict limit on the allowed clustering amplitude values and clustering offering a way to measure dark energy, we may hope that the Stage-IV surveys will be able to confidently exclude large regions of the extended parameter space.

\section*{Acknowledgements}

We would like to thank Daniel Farrow, and Martha Lippich
for their help and useful suggestions. 
This research was supported by the Excellence Cluster ORIGINS, 
which is funded by the Deutsche 
Forschungsgemeinschaft (DFG, German Research Foundation) under 
Germany's Excellence Strategy - EXC-2094 - 390783311.

AE is supported at the AIfA by an Argelander Fellowship.

JH has received funding from the European Union's Horizon 2020 research and innovation program under the Marie Sk\l{}odowska-Curie grant agreement No 101025187.

G.R. acknowledges support from the National Research Foundation of Korea (NRF) through Grant No. 2020R1A2C1005655 funded by the Korean Ministry of Education, Science and Technology (MoEST), and from the faculty research fund of Sejong University in 2022/2023.

Funding for the Sloan Digital Sky Survey IV has been provided by the Alfred P. Sloan Foundation, the U.S. Department of Energy Office of Science, and the Participating Institutions. SDSS-IV acknowledges
support and resources from the Center for High-Performance Computing at
the University of Utah. The SDSS web site is www.sdss.org.

\section*{Data Availability}

The clustering measurements from BOSS and eBOSS used in this analysis are publicly available 
 via the SDSS Science Archive Server (https://sas.sdss.org/).



\bibliographystyle{mnras}
\bibliography{main} 




\appendix

\section{Additional constraints}
\label{appx:A}

\begin{table*}
	\centering
	\caption{Marginalised posterior constraints (mean values with 68 per-cent confidence interval, for $\sum m_{\nu}$ - 95 per-cent confidence interval) derived from Planck CMB and the full 
	shape analysis of BOSS + eBOSS clustering measurements on their own, as well as in combination with each other and with Pantheon supernovae Ia measurements (SN). All of the models considered here vary dark energy equation of state parameter $w$, $w_a$CDM additionally allows a redshift evolution for $w$, $wK$CDM varies curvature and $w\nu$CDM varies neutrino mass sum $\sum m_{\nu}$. Note that for $wK$CDM the joint BOSS+eBOSS+Planck constraints should be interpreted bearing in mind that BOSS+eBOSS and Planck are discrepant in this parameter space. }
	\label{tab:constraints_apx}
	\begin{tabular}{c|c|ccc} 
		\hline
		\multicolumn{1}{c|}{} & \multicolumn{1}{c|}{} & \multicolumn{1}{c|}{BOSS+eBOSS} &  \multicolumn{1}{c|}{BOSS+eBOSS+Planck} & \multicolumn{1}{c|}{BOSS+eBOSS+Planck+SN} \\
		\hline
		\multirow{7}{3em}{$w$CDM}   & $\sigma_{8}$ & $0.798\pm 0.047$ & $0.828\pm 0.017$ & $0.818\pm 0.012$  \\
	                            	& $H_0$ & $69.8^{+3.1}_{-3.6}$ & $69.6^{+1.4}_{-1.6}$ & $68.62\pm 0.84$ \\
		& $\Omega_{\rm{m}}$ & $0.280^{+0.017}_{-0.021}$ & $0.295\pm 0.013$ & $0.3026\pm 0.0080$   \\
		& $\Omega_{\Lambda}$ & $0.720^{+0.021}_{-0.017}$ & $0.705\pm 0.013$ & $0.6974\pm 0.0080$ \\
		& $\omega_{\rm{m}}$ & $0.137^{+0.011}_{-0.013}$ & $0.1426\pm 0.0011$ & $0.1424\pm 0.0011$ \\
		& $n_{\rm{s}}$ & $0.990\pm 0.055$ & $0.9661\pm 0.0042$ & $0.9665\pm 0.0041 $ \\
		& $\rm{ln}10^{10}\rm{A}_{\rm{s}}$ & $3.02\pm 0.21$ & $3.043\pm 0.016$ & $3.044\pm 0.016$  \\
		\hline
		\multirow{7}{3em}{$w_a$CDM}  & $\sigma_{8}$ & $0.793\pm 0.045$  & $0.818^{+0.019}_{-0.022}$ & $0.822\pm 0.012$  \\
	                            	& $H_0$ & $70.1^{+3.7}_{-4.4}$ & $68.1^{+2.0}_{-2.8}$ & $68.72\pm 0.86$  \\
		& $\Omega_{\rm{m}}$ & $0.281^{+0.026}_{-0.030}$ & $0.309^{+0.024}_{-0.021}$ & $0.3025\pm 0.0081$  \\
		& $\Omega_{\Lambda}$ & $0.719^{+0.030}_{-0.026}$ & $0.691^{+0.021}_{-0.024}$ & $0.6975\pm 0.0081$ \\
		& $\omega_{\rm{m}}$ & $0.138\pm 0.012$ & $0.1428\pm 0.0012$ & $0.1427\pm 0.0011$ \\
		& $n_{\rm{s}}$ & $0.983\pm 0.054$ & $0.9656\pm 0.0042$ & $0.9658\pm 0.0041$ \\
		& $\rm{ln}10^{10}\rm{A}_{\rm{s}}$ & $3.00\pm 0.20$ & $3.042\pm 0.016$ & $3.042\pm 0.016$ \\
		\hline
		\multirow{7}{3em}{$wK$CDM}   & $\sigma_{8}$ & $0.770\pm 0.049 $ & $0.834\pm 0.019$ & $0.819\pm 0.012$  \\
	                            	& $H_0$ & $68.9^{+2.9}_{-3.4}$ & $69.7^{+1.4}_{-1.6}$ & $68.46\pm 0.91$ \\
		& $\Omega_{\rm{m}}$ & $0.292\pm 0.019$ & $0.292\pm 0.014 $ & $0.3032^{+0.0077}_{-0.0086}$   \\
		& $\Omega_{\Lambda}$ & $0.829\pm 0.073$ & $0.710\pm 0.015$ & $0.6982^{+0.0085}_{-0.0076}$ \\
		& $\omega_{\rm{m}}$ & $0.139^{+0.011}_{-0.013}$ & $0.1419\pm 0.0013$ & $0.1420\pm 0.0013$ \\
		& $n_{\rm{s}}$ & $0.975^{+0.060}_{-0.053}$ & $0.9679\pm 0.0046$ & $0.9676\pm 0.0045$  \\
		& $\rm{ln}10^{10}\rm{A}_{\rm{s}}$ & $2.80\pm 0.26$ & $3.041\pm 0.016$ & $3.043\pm 0.016$ \\
		\hline
		\multirow{7}{3em}{$w\nu$CDM}   & $\sigma_{8}$ & $0.795^{+0.041}_{-0.048}$ & $0.822\pm 0.018$  & $0.816^{+0.016}_{-0.013}$  \\
	                            	& $H_0$ & $70.3\pm 3.5$ & $70.0^{+1.5}_{-1.8}$ & $68.66\pm 0.85$ \\
		& $\Omega_{\rm{m}}$ & $0.288^{+0.019}_{-0.022}$ & $0.293\pm 0.013$  & $0.3028\pm 0.0083$  \\
		& $\Omega_{\Lambda}$ & $0.712^{+0.022}_{-0.019}$ & $0.707\pm 0.013$ & $0.6972\pm 0.0083$\\
		& $\omega_{\rm{m}}$ & $0.143^{+0.012}_{-0.014}$ & $0.1433^{+0.0013}_{-0.0016}$ & $0.1427\pm 0.0013$ \\
		& $n_{\rm{s}}$ & $1.066^{+0.070}_{-0.11}$ & $0.9659\pm 0.0039$ & $0.9665\pm 0.0041$ \\
		& $\rm{ln}10^{10}\rm{A}_{\rm{s}}$ & $3.18^{+0.23}_{-0.26}$ & $3.043\pm 0.016$ & $3.045\pm 0.016$ \\
		\hline
	\end{tabular}

\end{table*}

In this section, we present the constraints on parameters omitted in Table \ref{tab:constraints}, including constraints in the ``traditional" parameter space (Table \ref{tab:constraints_apx}). The physical parameters (i.e., not defined through $h$ units) constrained by our data include the physical matter density $\omega_{\rm m}$, the spectral index $n_{\rm s}$ and the (log) amplitude of initial density fluctuations $\rm{ln}10^{10}\rm{A}_{\rm{s}}$. For completeness, we include the traditional parameters $\sigma_8$ (linear density field variance as measured on the scale of $8h^{-1}$Mpc whose physical equivalent is $\sigma_{12}$), Hubble parameter $H_0$ and the relative densities of matter ($\Omega_{\rm m}$), dark energy ($\Omega_{\rm DE}$) and curvature ($\Omega_{\rm K}$). 

In general, we expect that the constraints on these parameters are degraded in comparison to their physical equivalents due to averaging over the posterior of $H_0$, which tends to be less well constrained in these extended cosmologies (this is most evident when comparing $\sigma_{12}$ with $\sigma_{8}$).

\bsp	
\label{lastpage}
\end{document}